# Realizing record-high transverse thermoelectric figure of merit at room temperature in artificially tilted multilayers based on high power factor NiFe alloy

Yebin Lee,[ab] Fuyuki Ando,*[a] Takamasa Hirai,[ac] Keisuke Hirata,[ac] Kota Hasegawa[ad] and Ken-ichi Uchida*[abc]

Transverse thermoelectric conversion using artificially tilted multilayers (ATMLs) offers a versatile device architecture that circumvents the structural limitations of conventional longitudinal thermoelectrics. However, achieving competitive room-temperature thermoelectric performance without an external magnetic field remains a critical challenge. Here, we report a record-high transverse thermoelectric figure of merit $z_{yx}T$ of 0.36 in $Ni_{50}Fe_{50}/Bi_{0.2}Sb_{1.8}Te_3$-based ATML at room temperature without an external magnetic field. Leveraging the longitudinal high power factor in a $Ni_{50}Fe_{50}$ alloy and the sharp contrast in electrical and thermal transport properties between *n*-type $Ni_{50}Fe_{50}$ and *p*-type $Bi_{0.2}Sb_{1.8}Te_3$, we engineer an anisotropic structure that simultaneously exploits high electrical conductivity, large transverse thermopower, and low thermal conductivity to maximize $z_{yx}T$ in ATML. Through the direct measurements of these thermoelectric transport parameters, we obtained $z_{yx}T$ of 0.36 in $Ni_{50}Fe_{50}/Bi_{0.2}Sb_{1.8}Te_3$-based ATML, which is in excellent agreement with the analytical prediction of 0.36 owing to the low interfacial electrical and thermal resistances at the $Ni_{50}Fe_{50}/Bi_{0.2}Sb_{1.8}Te_3$ junctions. These results pave the way for the practical implementation of transverse thermoelectric materials around room temperature.

## 1. Introduction

Thermoelectric conversion, which enables the direct interconversion between heat and electricity, is a promising technology for sustainable energy harvesting and solid-state cooling.[1] Conventional thermoelectric generators and refrigerators, based on the longitudinal Seebeck and Peltier effects, typically require complex $\pi$-type modules that consist of multiple pairs of *p*- and *n*-type semiconductors connected thermally in parallel and electrically in series. This complicated three-dimensional architecture is inherently limited by cumulative device-level performance degradation, including mechanical fragility and substantial efficiency losses originating from interfacial electrical and thermal resistances at the numerous junctions.[2] To overcome these structural limitations, transverse thermoelectric conversion has recently attracted considerable attention.[3,4] In transverse thermoelectrics, the charge and heat currents flow in mutually orthogonal directions, enabling a monolithic, junction-free, and versatile device architecture. This orthogonal geometry eliminates the need for complex junctions and metallic electrodes on the hot side, effectively mitigating the contact resistance and thermal degradation issues of conventional longitudinal modules, thereby expanding the applicability of thermoelectric devices for flexible and scalable thermal energy harvesting.[5]

As comprehensively summarized in a recent review by Adachi *et al.*,[6] transverse thermoelectric phenomena can be broadly classified based on their underlying symmetry breaking mechanisms. The first category is driven by time-reversal symmetry breaking; the representative phenomena in this category are the ordinary/anomalous Nernst effects (ONE/ANE).[7–11] While ONE can generate large transverse thermoelectric responses in Dirac or Weyl semimetals, it inevitably requires an external magnetic field, rendering it unsuitable for practical applications.[12–17] On the other hand, ANE in permanent magnets allows magnetic-field-free operation;[18,19] however, its transverse thermoelectric figure of merit $z_{yx}T$ is generally limited to $\sim 10^{-3}$ at room temperature, which is insufficient for practical use. The second category relies on structural symmetry breaking, which induces the off-diagonal Seebeck effect (ODSE).[20–22] While the goniopolar single crystal $Re_4Si_7$ exhibits exceptionally large $z_{yx}T$ of 0.7 at 980 K arising from microscale anisotropic structure,[23] its performance is limited to high temperatures and requires challenging single crystal growth. Alternatively, macroscale anisotropic structure can be engineered in artificially tilted multilayers (ATMLs) by alternately and obliquely stacking two constituent materials with high contrast in their transport properties. This geometrically tilted structure generates a substantial transverse thermopower through the anisotropic flow of charge and heat. To date, various ATML systems have been investigated to enhance $z_{yx}T$.[24–30] For instance, a recent report on $SmCo_5/Bi_{0.2}Sb_{1.8}Te_3$ (BST)-based ATML demonstrated $z_{yx}T$ of 0.20 at room temperature, achieved by minimizing interfacial electrical and thermal resistances through optimized anisotropic structure.[30] Furthermore, the scope of ATMLs has been expanded by superposing magneto-thermoelectric effects including ONE and ANE with ODSE.[31–34] Such hybridization strategies, leveraging the additive contribution of ANE and anisotropic thermal conductivity in the constituent material in ATML, have successfully pushed the performance to $z_{yx}T \approx 0.30$ at room temperature in in-plane magnetized $SmCo_5$/BST-based ATML.[33] However, further advances in material design are is required for realizing high-performance transverse thermoelectric conversion toward practical thermal management applications around room temperature.

In this work, we report a record-high transverse thermoelectric performance solely by ODSE in ATML comprising BST and Ni-based-alloy slabs. Ni-based alloys have recently attracted significant attention as metallic thermoelectric materials owing to their ultrahigh power factors.[35–38] Such materials are particularly attractive candidates for constructing ATMLs because they show a relatively large negative Seebeck coefficient and a large contrast in electrical and thermal transport properties against conventional *p*-type thermoelectric materials, such as BST. Among them, we selected a $Ni_{50}Fe_{50}$ (NiFe) alloy that exhibits a higher power factor than pure Ni.[35,37,39,40] By combining NiFe with *p*-type BST, we achieved a transverse figure of merit of $z_{yx}T = 0.36$ at room temperature, which closely matches the analytical calculations predicting $z_{yx}T = 0.36$. Furthermore, all transport parameters relevant to $z_{yx}T$ together with the electrical and thermal interfacial resistances were experimentally determined. These results demonstrate the significant potential of NiFe/BST-based ATML for highly efficient thermal energy harvesting and electronic cooling.

[a] National Institute for Materials Science, Tsukuba 305-0047, Japan.
E-mail: ANDO.Fuyuki@nims.go.jp; UCHIDA.Kenichi@nims.go.jp
[b] Graduate School of Science and Technology, University of Tsukuba, Tsukuba 305-8573, Japan.
[c] Department of Advanced Materials Science, Graduate School of Frontier Sciences, The University of Tokyo, Kashiwa 277-8561, Japan.
[d] Resonac Corporation, Tsukuba 300-4247, Japan

## 2. Results and discussion

### 2.1 Structural design and calculation of transverse thermoelectric properties

To maximize the transverse thermoelectric performance driven by ODSE, we designed ATML as schematically illustrated in Fig. 1a. In this configuration, the alternate and oblique stacking of two constituent materials with highly contrasting electrical and thermal properties skews the pathways of charge and heat flows. When a temperature gradient $\nabla_x T$ along the $x$ direction is applied, the anisotropic transport within the tilted boundaries induces an electric field $\mathbf{E}_y$ along the $y$ direction.[41–45] For clarity, we define the $x$ and $y$ directions in this work as the directions of the applied temperature gradient and generated electric field, respectively; this convention differs from that used in previous studies that address both heat-to-electricity and its reciprocal electricity-to-heat conversion.[30–33]

The fundamental requirement for maximizing ODSE in ATML is the selection of an optimal pair of materials with opposite Seebeck coefficients $S$ and a suitable contrast in their electrical conductivity $\sigma$ and thermal conductivity $\kappa$. In this study, we selected $p$-type BST as the thermoelectric counterpart to the NiFe alloy because its large positive Seebeck coefficient and low conductivities complement the large negative Seebeck coefficient and high conductivities of NiFe. Fig. 1b shows the experimentally measured transport properties of the constituent NiFe and BST slabs, confirming the combination of opposite Seebeck coefficients and a large contrast in transport properties required for generating a large transverse thermopower in ATML.

Based on the measured transport parameters of the individual NiFe and BST layers, we calculated transverse thermoelectric properties of NiFe/BST-based ATML using the analytical methods in previous studies.[44,45] First, $\sigma$, $\kappa$, and $S$ of NiFe and BST in the directions parallel ($\sigma_\parallel$, $\kappa_\parallel$, and $S_\parallel$) and perpendicular ($\sigma_\perp$, $\kappa_\perp$, and $S_\perp$) to the stacking plane are determined by the thickness ratio of the NiFe layer $t = d_{\mathrm{NiFe}}/(d_{\mathrm{NiFe}}+d_{\mathrm{BST}})$ with $d_{\mathrm{NiFe}}$ and $d_{\mathrm{BST}}$ respectively being the thickness of the NiFe and BST layers as

$$\sigma_\parallel = t\sigma_{\mathrm{NiFe}} + (1-t)\sigma_{\mathrm{BST}}$$
$$\sigma_\perp = \frac{\sigma_{\mathrm{NiFe}}\sigma_{\mathrm{BST}}}{(1-t)\sigma_{\mathrm{NiFe}} + t\sigma_{\mathrm{BST}}} \quad (1)$$

$$\kappa_\parallel = t\kappa_{\mathrm{NiFe}} + (1-t)\kappa_{\mathrm{BST}}$$
$$\kappa_\perp = \frac{\kappa_{\mathrm{NiFe}}\kappa_{\mathrm{BST}}}{(1-t)\kappa_{\mathrm{NiFe}} + t\kappa_{\mathrm{BST}}} \quad (2)$$

$$S_\parallel = \frac{t\sigma_{\mathrm{NiFe}}S_{\mathrm{NiFe}} + (1-t)\sigma_{\mathrm{BST}}S_{\mathrm{BST}}}{t\sigma_{\mathrm{NiFe}} + (1-t)\sigma_{\mathrm{BST}}}$$
$$S_\perp = \frac{t\kappa_{\mathrm{BST}}S_{\mathrm{NiFe}} + (1-t)\kappa_{\mathrm{NiFe}}S_{\mathrm{BST}}}{t\kappa_{\mathrm{BST}} + (1-t)\kappa_{\mathrm{NiFe}}} \quad (3)$$

Subsequently, the transverse transport coefficients of ATML were analytically calculated as functions of the tilt angle $\theta$ described in Fig. 1a:

$$\sigma_{yy} = \frac{\sigma_\parallel\sigma_\perp}{\sigma_\parallel \sin^2\theta + \sigma_\perp \cos^2\theta} \quad (4)$$

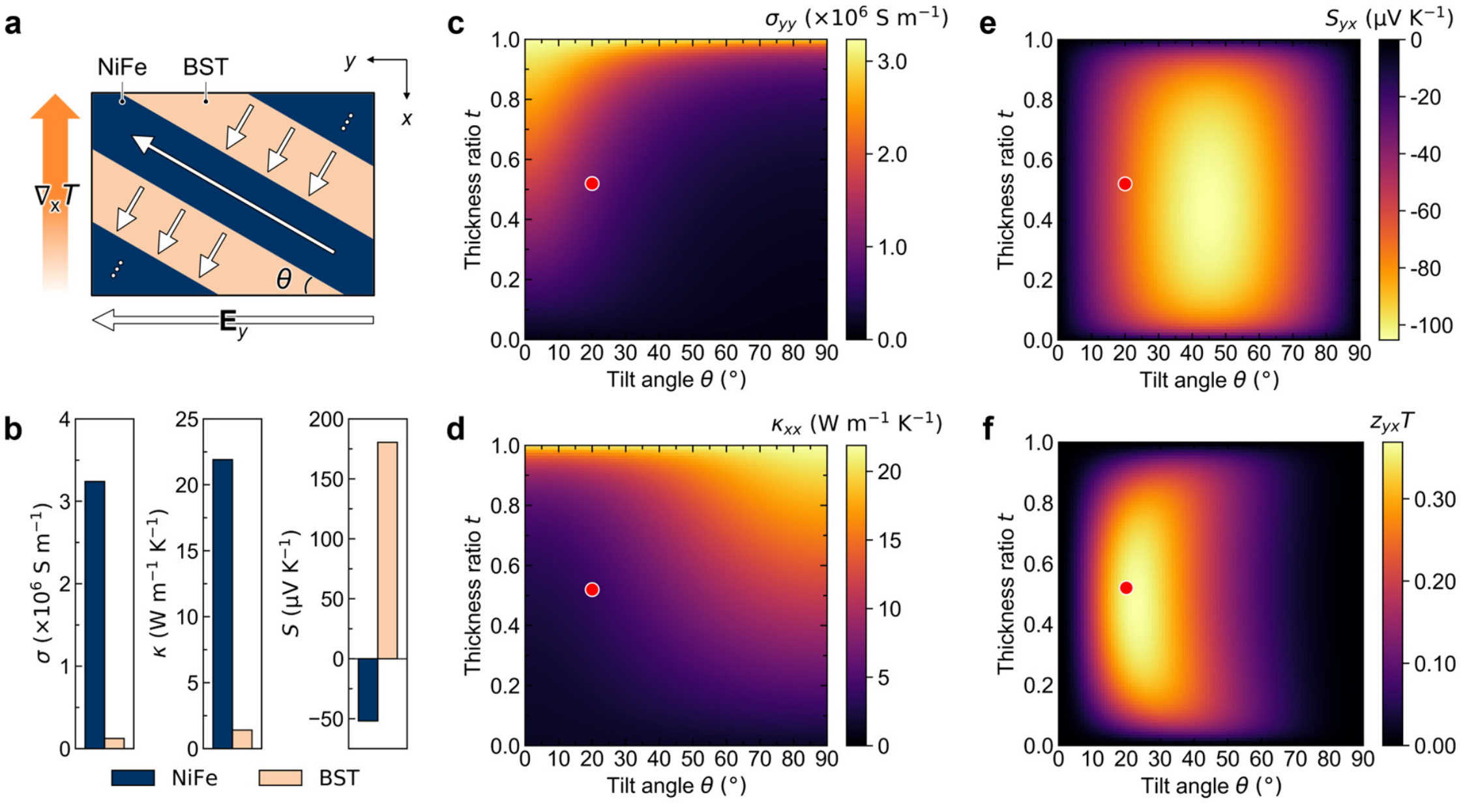

**Fig. 1** Design and calculated performance of the $Ni_{50}Fe_{50}$ (NiFe)/$Bi_{0.2}Sb_{1.8}Te_3$ (BST)-based artificially tilted multilayer (ATML). (a) Schematic illustration of the transverse thermoelectric conversion in ATML composed of alternately and obliquely stacked NiFe and BST layers. A temperature gradient $\nabla_x T$ along the $x$ direction induces an electric field $\mathbf{E}_y$ along the $y$ direction driven by the off-diagonal Seebeck effect. The white arrows inside ATML represent the pathways of the charge current. (b) Experimentally measured electrical conductivity $\sigma$, thermal conductivity $\kappa$, and Seebeck coefficient $S$ of the individual NiFe and BST slabs at room temperature. (c)–(f) Calculated two-dimensional contour maps of the transverse thermoelectric properties for NiFe/BST-based ATML at 300 K as functions of the tilt angle $\theta$ and the thickness ratio of the NiFe layer $t$: (c) electrical conductivity along the $y$ direction $\sigma_{yy}$, (d) thermal conductivity along the $x$ direction $\kappa_{xx}$, (e) transverse thermopower $S_{yx}$, and (f) transverse thermoelectric figure of merit $z_{yx}T$. The red circles denote the optimized geometric parameters ($\theta$ = 20° and $t$ = 0.52) adopted for the actual ATML fabricated in this study.

$$\kappa_{xx} = \kappa_{||}\sin^2\theta + \kappa_{\perp}\cos^2\theta \quad (5)$$

$$S_{yx} = (S_{||} - S_{\perp}) \sin\theta \cos\theta \quad (6)$$

Here, $S_{yx}$ is the transverse thermopower, $\sigma_{yy}$ is the electrical conductivity along the $y$ direction, and $\kappa_{xx}$ is the thermal conductivity along the $x$ direction in ATML. Then, the transverse figure of merit $z_{yx}T$ is given by

$$z_{yx}T = \frac{{S_{yx}}^2\sigma_{yy}}{\kappa_{xx}}T \quad (7)$$

Fig. 1c–f display the two-dimensional contour maps of the calculated transverse thermoelectric properties at 300 K. The analytical calculation clearly visualizes the strong geometric dependence of ATML performance. Notably, the combination of the prominent negative Seebeck coefficient of NiFe and the contrast in $\sigma$ and $\kappa$ between NiFe and BST leads to exceptionally high $z_{yx}T$ even at room temperature. The calculated maximum $z_{yx}T$ value reaches 0.36 at $T$ = 300 K around the optimal geometric conditions. Guided by these calculated contour maps, we fabricated NiFe/BST-based ATML with $\theta = 20°$ and $t = 0.52$ using spark plasma sintering (SPS, see details in Experimental section).

## 2.2 Microstructural characterization and interfacial transport properties

To achieve high thermoelectric performance in ATMLs, minimizing the contribution of interfacial electrical and thermal resistances is essential, as the multitude of heterogeneous interfaces in ATMLs inherently introduces electrical and thermal contact resistances, which have frequently been identified as a primary cause of significant performance degradation in previous thermoelectric materials.[46–49] Therefore, we first investigated the microstructural features of the NiFe/BST junctions to verify the structural integrity of the fabricated multilayer and identify possible interfacial reaction layers formed during SPS processing.

Fig. 2a displays the cross-sectional scanning electron microscopy (SEM) image with the corresponding energy-dispersive X-ray spectroscopy (EDS) elemental mapping and line profile of the NiFe/BST multilayer at low magnification. The constituent NiFe and BST layers are alternately and densely stacked without any observable macroscopic voids or cracks, confirming the successful consolidation via SPS process. To further investigate the interfacial reaction, the high-magnification observation across the NiFe/BST interface was performed, as shown in Fig. 2b. The distributions of Ni, Fe, Bi, Sb, and Te indicate that the interfaces remain clearly distinguishable without catastrophic bulk alloying. Notably, a finite elemental diffusion layer of approximately 50 μm is observed at the interfaces, where the FeTe and NiTeSb alloy regions appear sequentially.[29,50] This stratified interfacial reaction is primarily driven by the differences in the elemental diffusion rates and chemical reactivities among the constituent elements during the high-temperature SPS process. While such a diffusion layer serves as a robust adhesive bond that ensures mechanical integrity of the multilayer structure, its presence raises the question of whether it introduces non-negligible parasitic resistances that could degrade the macroscopic transport contrast required for ODSE.

To quantitatively address this, we investigated the interfacial transport properties across the NiFe/BST junctions. The spatial distribution of the electrical resistance was evaluated by measuring the resistance while scanning a voltage probe along the stacking direction (Fig. 3a). As shown in Fig. 3b and its magnified view of the interface in Fig. 3c, the resistance profile exhibits a step-like behavior reflecting the contrasting electrical resistivities of the NiFe and BST layers. The bulk resistivities estimated from the slopes of the linear fit at the NiFe and BST regions were consistent with those independently obtained for the individual NiFe and BST slabs shown in Fig. 1b. By extrapolating the linear fits of the NiFe and BST regions to the interfaces, the averaged interfacial electrical resistance across several junctions was estimated to be 0.9±1.0 μΩ cm$^2$ (red arrows in Fig. 3c). In comparison, the bulk resistance of the constituent NiFe and BST layers was 37±9 μΩ cm$^2$. This result demonstrates that the interfacial electrical resistance accounts for a small fraction (approximately 5±5%) of the bulk electrical resistance, indicating a minimized impact on macroscopic electrical transport.

Next, the interfacial thermal resistance was assessed by analyzing the temperature distribution under a thermally nonequilibrium state across the multilayers. As depicted in Fig.

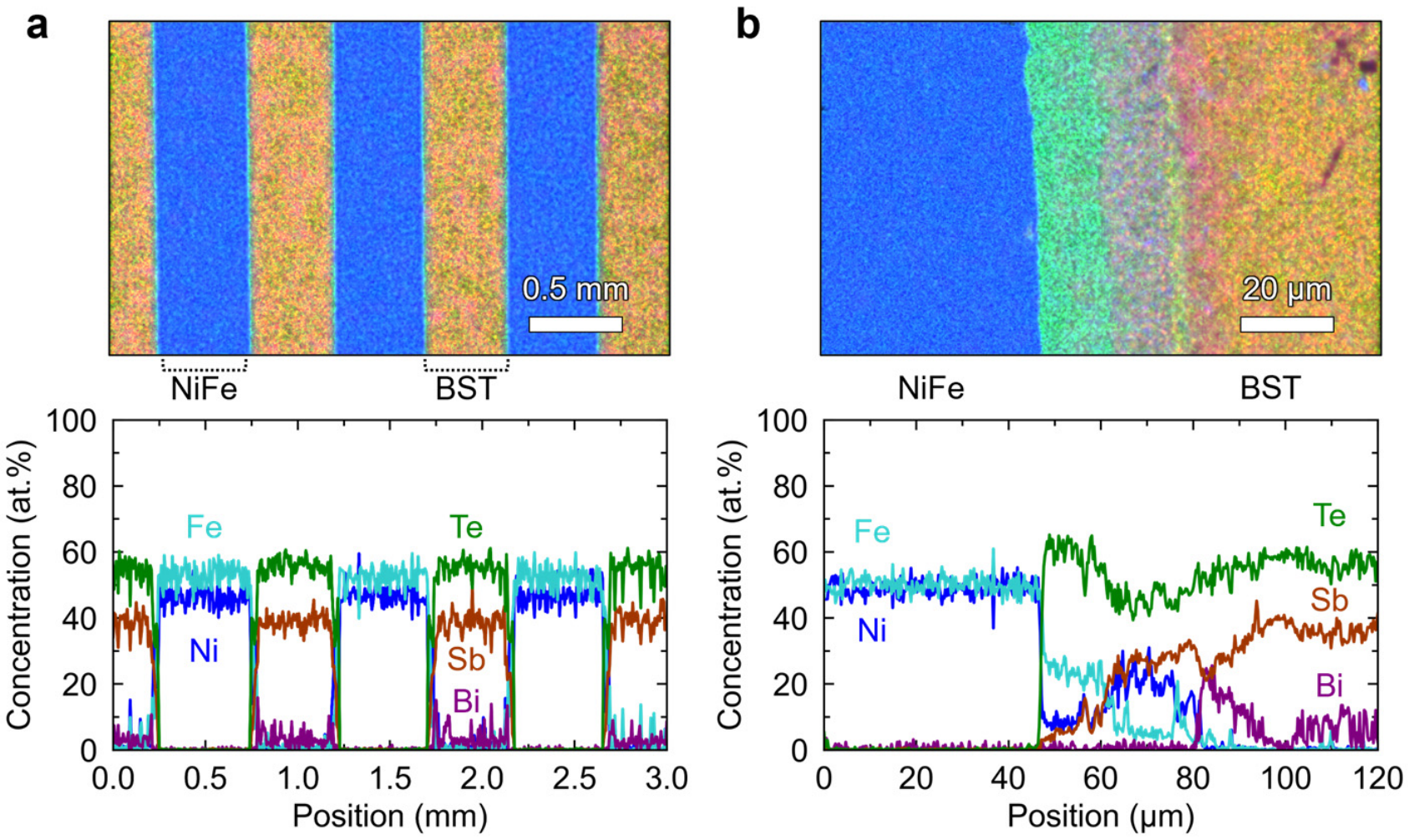

**Fig. 2** Structural and composition characterization of the NiFe/BST multilayer. Cross-sectional scanning electron microscopy with energy-dispersive X-ray spectroscopy images for the multilayer at (a) low and (b) high magnifications. Line profiles of the atomic ratios of Ni, Fe, Bi, Sb, and Te across the stacking direction corresponding to each mapping image are shown below.

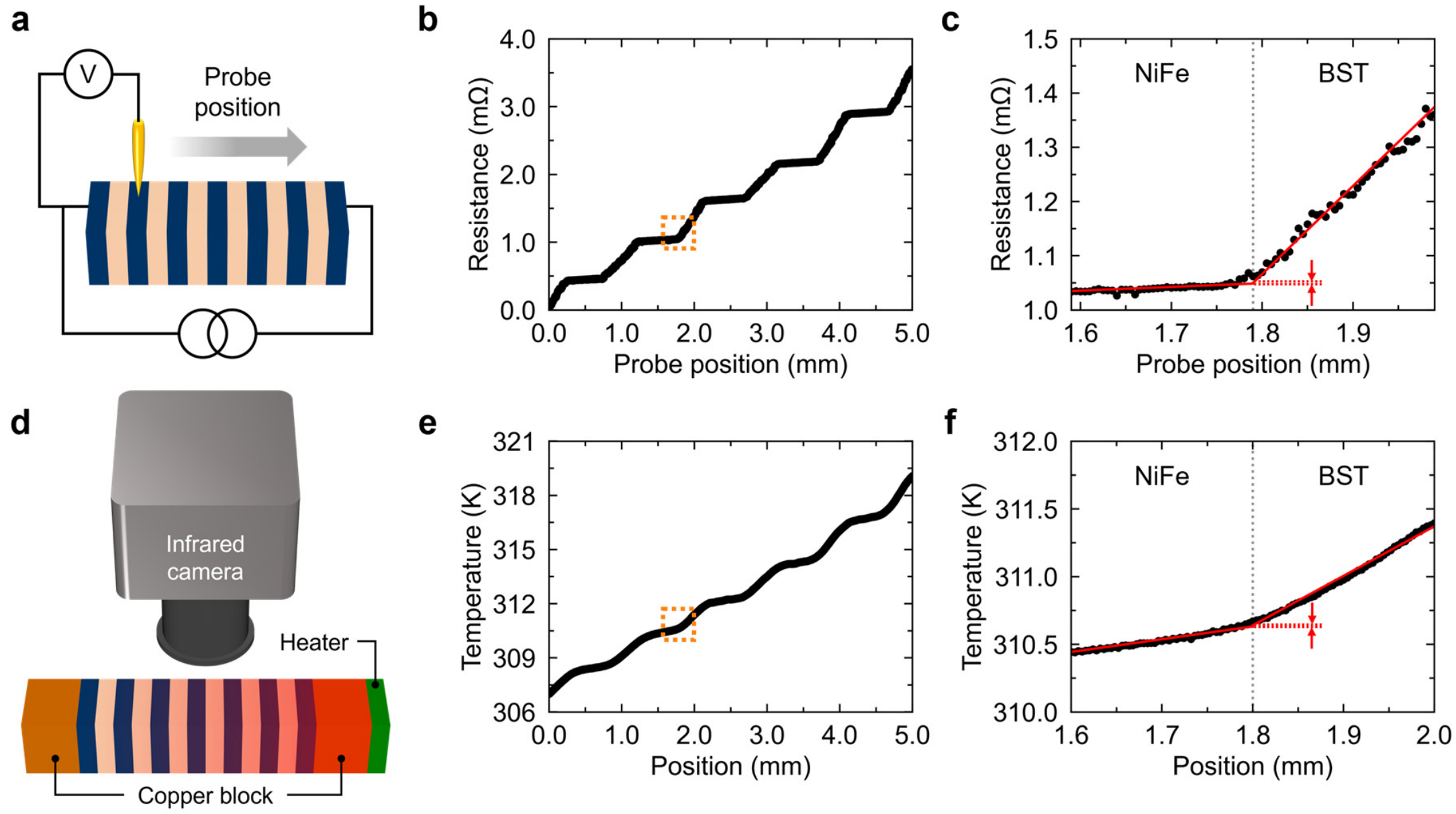

**Fig. 3** Evaluation of the interfacial transport properties of the NiFe/BST multilayer. (a) Schematic of the spatial electrical resistance measurement obtained by moving a contact probe along the stacking direction. (b) Position-dependent electrical resistance profile exhibiting a step-like behavior across the multilayer. (c) Magnified view of the electrical resistance profile at the NiFe/BST interface indicated by an orange dotted square in (b). The gray dotted line is the approximate position of the NiFe/BST interface, where a gap indicated by red arrows corresponds to the interfacial electrical resistance. (d) Schematic setup for mapping the spatial temperature distribution across the multilayer utilizing an infrared camera, where the sample is sandwiched between copper blocks attaching a resistance heater to apply a steady-state heat flow. (e) Position-dependent temperature profile across the multilayer. (f) Magnified view of the temperature profile at the interface in (e). A gap indicated by red arrows shows the temperature drop due to the interfacial thermal resistance.

3d, the sample was bridged between copper blocks attaching a resistance heater to apply a steady-state heat flow, and an infrared camera was utilized to map the temperature profile. The position-dependent temperature curve (Fig. 3e) and its magnified view of the interface (Fig. 3f) reveal a continuous temperature gradient with a minimal temperature jump at the interface. By evaluating the temperature gaps at multiple interfaces using the one-dimensional heat diffusion equation, the averaged interfacial thermal resistance was characterized to be 1.6±1.2 μW$^{-1}$ m$^2$ K, whereas the bulk thermal resistance of the NiFe and BST layers derived from the slopes of the linear fits was 46.8±11.2 μW$^{-1}$ m$^2$ K. This low interfacial thermal resistance, constituting about 7±5% of the bulk thermal resistance, proves that heat transport is mainly governed by the intrinsic properties of the constituent materials rather than being bottlenecked at the interfaces.

Consequently, the well-defined interfaces with minimized parasitic resistances ensure nearly ideal electrical and thermal transport in NiFe/BST-based ATML, thereby supporting the high $z_{yx}T$ values presented in the following section.

### 2.3 Quantitative evaluation of transverse thermoelectric performance

To validate the transverse thermoelectric performance of NiFe/BST-based ATML, we directly measured all the transport parameters, $S_{yx}$, $\sigma_{yy}$, and $\kappa_{xx}$, at room temperature. First, $S_{yx}$ and $\sigma_{yy}$ of ATML were evaluated following the procedures established in previous study.[30] $S_{yx}$ was characterized by measuring the electric field $E_y$ generated along the $y$ direction while applying $\nabla_x T$ along the $x$ direction (Fig. 4a). Specifically, $\nabla_x T$ was determined from the temperature distribution within the area indicated by the white box in Fig. 4a, while $E_y$ was obtained by measuring the voltage between the voltage probes represented by the two black dots. The voltage probes are slightly offset in the $y$ direction because they need to be attached on the NiFe surfaces, which ensures the Ohmic contact. We calibrate the transverse thermoelectric voltage by subtracting the longitudinal Seebeck contribution due to this position offset. The measured voltages are expressed as:

$$V_x = l_x E_x + \delta_y E_y \tag{8}$$

$$V_y = l_y E_y + \delta_x E_x \tag{9}$$

where $V_x$ ($V_y$) is the measured voltage between the probe pair aligned along the $x$ ($y$) direction, $l_x$ ($l_y$) denotes the $x$ ($y$) distance between the corresponding voltage probes, and $\delta_y$ ($\delta_x$) represents the position offset along the $y$ ($x$) direction. $E_y$ was then obtained by solving these simultaneous equations. As shown in Fig. 4b, the magnitude of $E_y$ scales linearly with $\nabla_x T$, yielding measured $S_{yx}$ of −68.4±0.2 μV K$^{-1}$. $\sigma_{yy}$ was determined to be 1.02 × 10$^6$ S m$^{-1}$ using a four-terminal method, which is consistent with the calculated value of 1.01 × 10$^6$ S m$^{-1}$. Moreover, unlike previous ATML studies that relied on calculated $\kappa_{xx}$ values, $\kappa_{xx}$ of our ATML was directly measured by the steady-state method, with a one-dimensional heat flux applied across ATML, giving $\kappa_{xx}$ = 4.00 W m$^{-1}$ K$^{-1}$, in good agreement with the calculated value of 3.83 W m$^{-1}$ K$^{-1}$.

The measured $S_{yx}$ value is in excellent agreement with the value (−67.3 μV K$^{-1}$) calculated under the ideal isothermal condition. As explained by Ando *et al.*,[51] thermal boundary condition during the actual measurement can affect the transport behavior. Fig. 4c shows the $\theta$ dependence of the adiabatic transverse thermopower $S_{yx}^*$, isothermal transverse thermopower $S_{yx}$, and longitudinal thermopower along the $y$ direction $S_{yy}$ for NiFe/BST-based ATML, calculated based on the analytical model proposed by Ando *et al.* The off-diagonal thermal conduction (ODTC) inherent to ATMLs redirects the heat flow. Under the nearly adiabatic condition in the experimental setup, ODTC induces a temperature gradient along the $y$ direction $\nabla_y T$. Consequently, $\nabla_y T$ induces a voltage by $S_{yy}$ that counteracts the ODSE-induced voltage around $\theta$ = 20° as depicted in Fig. 4c, giving $S_{yx}^* = -62.0\ \mu\mathrm{V\ K^{-1}}$. Nevertheless, the close match of the measured $S_{yx}$ value to the calculated one suggests that thermal boundary condition effects cause no significant deviation in the present measurement.

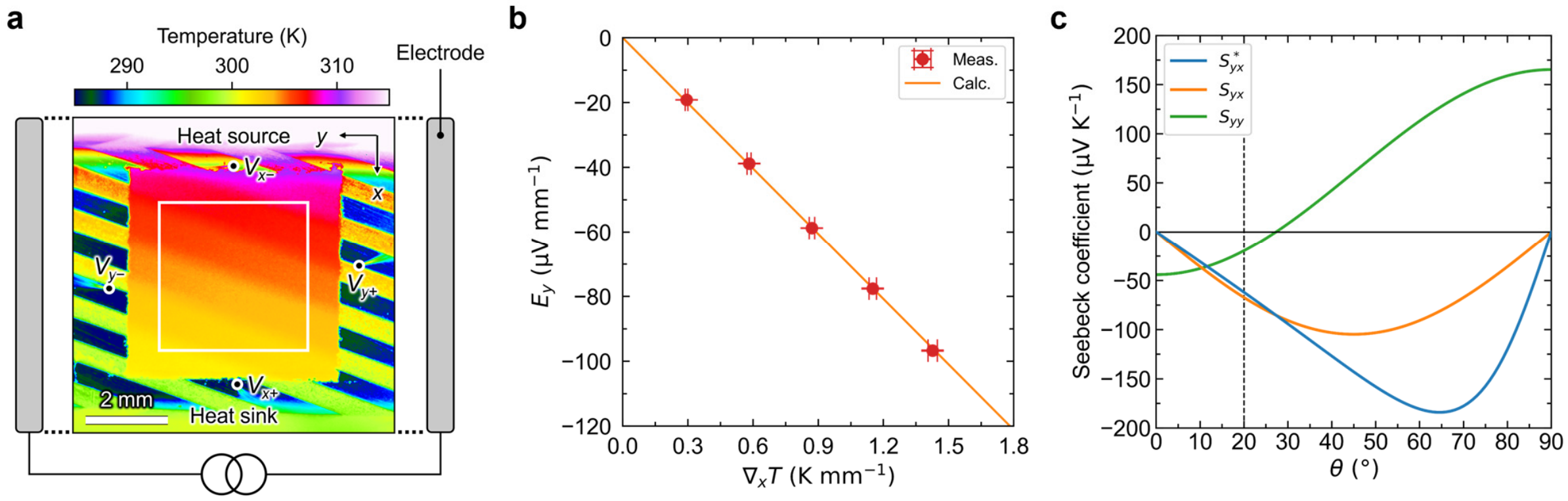

**Fig. 4** Characterization of the transverse thermopower of NiFe/BST-based ATML. (a) Infrared thermal image of NiFe/BST-based ATML under $\nabla_x T$ along the $x$ direction. $\nabla_x T$ was determined from the temperature distribution within the area indicated by the white box, while the electric field $E_y$ was obtained by measuring the voltage between the two black dots along the $y$ direction. (b) Measured (Meas.) and calculated (Calc.) $E_y$ as a function of the applied $\nabla_x T$. (c) $\theta$ dependence of the isothermal transverse thermopower $S_{yx}$, adiabatic transverse thermopower $S_{yx}^*$, and longitudinal thermopower $S_{yy}$ for NiFe/BST-based ATML. The vertical dashed line indicates $\theta$ = 20° of the synthesized ATML.

By incorporating these directly measured properties, we obtained experimentally determined $z_{yx}T$ of 0.36 at room temperature, in reasonable agreement with analytical prediction of 0.36. As benchmarked in Fig. 5, this value represents record-high $z_{yx}T$ among room-temperature ATML systems reported to date. It is worth noting that this comparison is made under zero magnetic field, which is crucial for practical applications (note that in the results in ref. 33, ANE is hybridized in the absence of an external magnetic field owing to the remanent magnetization of $SmCo_5$). Our NiFe/BST-based ATML exhibits a superior performance, outperforming previously reported ATML systems combining BST with various counterparts such as $SmCo_5$, Ni, $n$-type $Bi_2Te_{2.7}Se_{0.3}$ (BTS) and Co.[27–30,33] We note again that for these previously reported ATML systems, the calculated $\kappa_{xx}$ values were used to estimate $z_{yx}T$. In contrast, for our NiFe/BST-based ATML, all transport parameters including $\kappa_{xx}$ were directly and experimentally determined. Accordingly, realizing such high $z_{yx}T$ at zero-field highlights the superiority of NiFe/BST-based ATML and its promise as a platform for room-temperature transverse thermoelectric applications.

## 3. Conclusions

In summary, we experimentally demonstrated a record-high transverse thermoelectric figure of merit of 0.36 at room temperature without an external magnetic field in NiFe/BST-based ATML. By combining the remarkably large negative Seebeck coefficient of NiFe with suitable contrasts in electrical and thermal transport properties between NiFe and $p$-type BST and by adopting optimized geometric parameters, we maximized the overall transverse thermoelectric performance. Spatially resolved characterization confirmed small interfacial electrical and thermal resistances at the NiFe/BST junctions and the direct measurement of all transport coefficients provided a rigorously complete performance assessment. The measured $z_{yx}T$ value of 0.36 is well consistent with the analytically calculated value of 0.36 and is achieved at zero magnetic field, which is an essential prerequisite for practical deployment. A key challenge for the future is to systematically characterize the temperature dependence of the transport coefficients. This will clarify whether it is possible to achieve a further increase beyond the room-temperature $z_{yx}T$ value, thereby guiding the design of ATMLs for waste heat recovery and solid-state cooling applications.

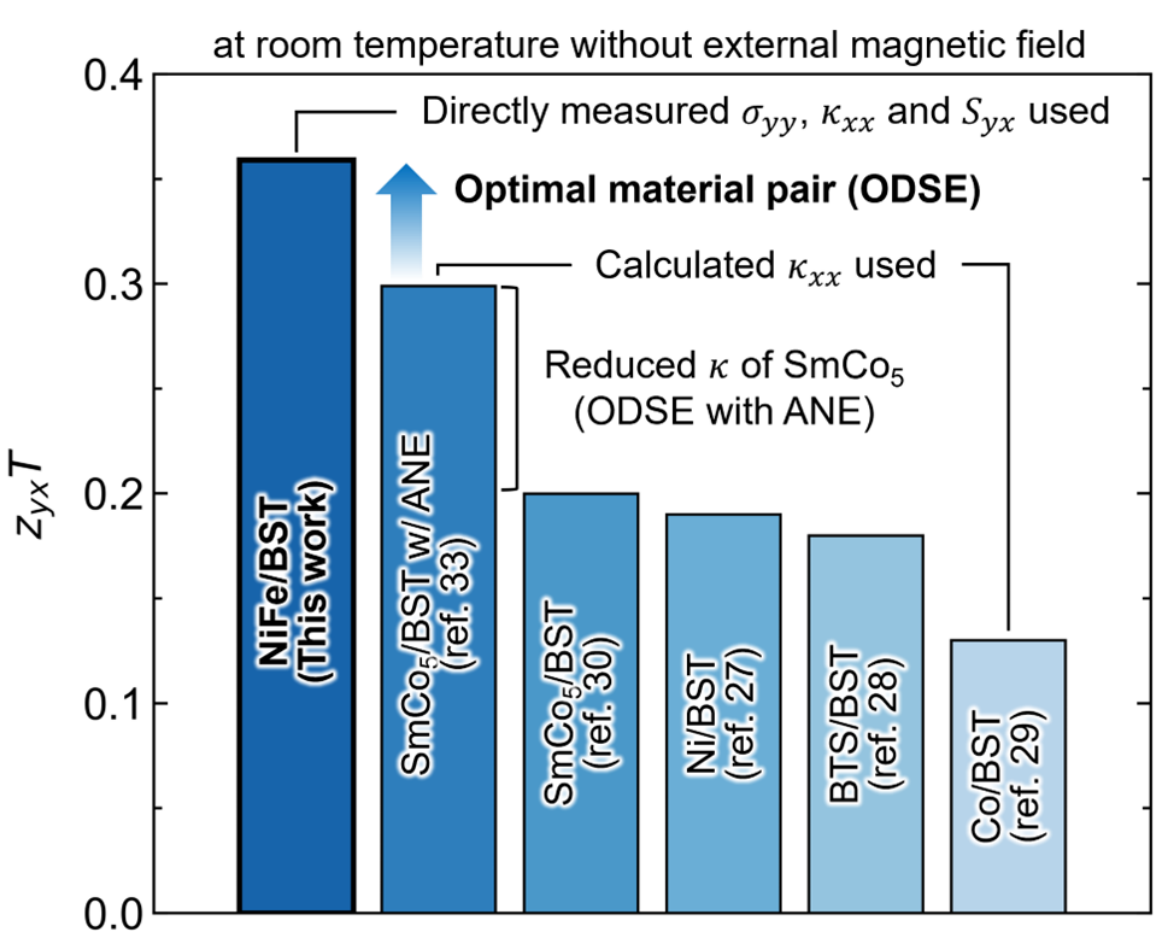

**Fig. 5** Comparison of the measured and calculated $z_{yx}T$ values at room temperature for various ATML systems combining BST with various counterparts such as NiFe, $SmCo_5$, Ni, $n$-type $Bi_2Te_{2.7}Se_{0.3}$ (BTS) and Co. Previous studies on ATMLs combined with $SmCo_5$, Ni, BTS and Co estimated $z_{yx}T$ using calculated $\kappa_{xx}$ values. The leftmost dark blue bar denotes experimental $z_{yx}T$ of 0.36 of NiFe/BST-based ATML obtained in this work. The remaining bars represent previously reported high-performance ATML systems, including those hybridized with the anomalous Nernst effect (ANE). All systems operate under zero magnetic field.

## 4. Experimental section

### 4.1. Sample preparation

The NiFe/BST multilayers were fabricated via SPS process. The polycrystalline NiFe rod, prepared by a melting method, with a diameter of 20 mm (Kojundo Chemical Laboratory Co., Ltd.) was sliced into 0.5-mm-thick disks using wire electrical discharge machining, followed by mechanical polishing. The NiFe disks and BST powders with 99.9% purity (Toshima Manufacturing Co., Ltd.) were alternately stacked and bonded by SPS under 30 MPa at 500 °C for 30 min in a vacuum. After the consolidation, the thicknesses of the NiFe and BST layers in the multilayer block were determined to be 0.51±0.01 mm and 0.47±0.02 mm, respectively. The sintered multilayer block was then cut into rectangular shapes using a diamond wire saw to obtain the ATML devices with $\theta = 20\pm0.2°$.

### 4.2. Microstructural and interfacial characterizations

The cross-sectional morphologies and elemental distributions of the NiFe/BST multilayers were characterized using SEM equipped with EDS (CrossBeam 1540EsB, Carl Zeiss AG).

To evaluate the interfacial transport properties, the NiFe/BST multilayer with $\theta = 90°$ and dimensions of 11.1 × 3.1 × 1.7 mm was prepared. The spatial distribution of the electrical resistance across the NiFe/BST interfaces was measured using a resistance distribution measuring instrument (M808, Mottainai Energy Co., Ltd.) by applying a charge current of 50 mA and moving a contact probe at 5 μm intervals. Furthermore, the interfacial thermal resistance was evaluated by applying a steady-state heat flow across the same $\theta = 90°$ sample, where the hot-side temperature was maintained at 330 K and the cold side of the sample was fixed at a heat bath. To accurately map the position-dependent temperature distribution, the top surface of the sample was coated with an insulating black ink having an emissivity of >0.94 (JSC-3, JAPANSENSOR Corp.) and thermal images were captured using an infrared camera (ImageIR 9450, InfraTec GmbH).

### 4.3. Thermoelectric property measurements

For the individual NiFe and BST slabs, $\sigma$ and $S$ were measured using the Seebeck Coefficient/Electric Resistance Measurement System (ZEM-3, ADVANCE RIKO, Inc.). The dimensions of the specimens used for the measurements were 3.0 × 12.1 × 1.0 mm for NiFe and 3.1 × 12.0 × 1.0 mm for BST, where the temperature gradient and charge current were applied along the longest direction of the samples. For the ATML sample, $\sigma_{yy}$ and $S_{yx}$ were evaluated using an ATML slab with dimensions of 12.3 × 8.0 × 1.1 mm. To apply a charge current, the 8.0 × 1.1 mm surfaces of the ATML slab were fully covered with Cerasolzer #297 (Komura-Tech Co., Ltd) using an ultrasonic soldering technique to form electrodes. Meanwhile, for accurate voltage and resistance probing, Al-1%Si wires were attached to the top surface of the sample. The $\sigma_{yy}$ was measured using a battery internal resistance tester (BT3562, HIOKI E.E. Corp.) by applying an alternating charge current with an amplitude of 10 mA. $\nabla_x T$ during the $S_{yx}$ measurement was determined utilizing the infrared camera (ELITE, DCG Systems G.K.), as described in Section 4.2. $\kappa$ of the individual NiFe and BST bulks was directly evaluated using a steady-state thermal conductivity measurement system (SS-H40, Bethel Co., Ltd.) with the hot-side temperature set to 330 K. The specimen dimensions were 12.1 × 12.1 × 12.0 mm for NiFe and 12.1 × 2.9 × 12.0 mm for BST, where the temperature gradient was applied along the 12.1 and 2.9 mm directions of the NiFe and BST slabs respectively. For the ATML sample, $\kappa_{xx}$ was measured for a sample with dimensions of 12.1× 8.1 × 12.1 mm using the same measurement system, where the temperature gradient was applied along the 8.1 mm direction.

## Author contributions

Y. L. and F. A. fabricated the samples and carried out the overall experimental investigations and characterization. Y. L. and T. H. measured the thermoelectric properties. Y. L., K. Hirata, and K. U. conducted the thermal conductivity measurements. Y. L. and K. Hasegawa performed measurements of interfacial thermal resistance. F. A. and K. U. designed and supervised the project. Y. L. analyzed the data and wrote the original manuscript. All authors discussed the results and contributed to the writing and revision of the manuscript.

## Conflicts of interest

There are no conflicts to declare.

## Data availability

The data that support the findings of this study are available from the corresponding author upon reasonable request.

## Acknowledgements

The authors thank N. Takahara, Y. Takano, K. Suzuki, and M. Isomura for technical supports. This work was supported by ERATO "Magnetic Thermal Management Materials" (No. JPMJER2201) from Japan Science and Technology Agency (JST) and Grant-in-Aid for Early-Career Scientists (KAKENHI) (No. 24K17610) from Japan Society for the Promotion of Science (JSPS).